\documentclass[%
 aip,% 
 jcp,% 
 amsmath,amssymb,% 
 floatfix,% 
 reprint,% 
 author-numerical,%
]{revtex4-1}

\usepackage{amsmath}
\usepackage{amssymb}
\usepackage{graphicx}
\usepackage{mathrsfs}
\usepackage{setspace}
\usepackage{bm} 

\newcommand{\kt}{k_B T}
\newcommand{\de}{\mathrm{d}}

\begin{document}

% Title
\title{Theory and simulations of toroidal and rod-like structures in single-molecule DNA condensation}

\author{Ruggero Cortini}
\email{cortini@lptl.jussieu.fr}

\author{Bertrand R.~Car\'e}
\email{care@lptl.jussieu.fr}

\author{Jean-Marc Victor}
\email{victor@lptl.jussieu.fr}

\author{Maria Barbi}
\email{barbi@lptl.jussieu.fr}
\affiliation{Laboratoire de Physique Th\'eorique de la Mati\`ere Condens\'ee, UMR
  7600, Universit\'e Pierre et Marie Curie, Sorbonne Universit\'e, 4 place
Jussieu, 75252 Cedex 05, Paris, France}

% Abstract
\begin{abstract}

DNA condensation by multivalent cations plays a crucial role in genome packaging
in viruses and sperm heads, and has been extensively studied using
single-molecule experimental methods. In those experiments, the values of the
critical condensation forces have been used to estimate the amplitude of the
attractive DNA-DNA interactions. Here, to describe these experiments, we
developed an analytical model and a rigid body Langevin dynamics assay to
investigate the behavior of a polymer with self-interactions, in the presence of
a traction force applied at its extremities. We model self-interactions using a
pairwise attractive potential, thereby treating the counterions implicitly. The
analytical model allows to accurately predict the equilibrium structures of
toroidal and rod-like condensed structures, and the dependence of the critical
condensation force on the DNA length. We find that the critical condensation
force depends strongly on the length of the DNA, and finite-size effects are
important for molecules of length up to 10$^5\mu\mathrm{m}$. Our Langevin
dynamics simulations show that the force-extension behavior of the rod-like
structures is very different from the toroidal ones, so that their presence in
experiments should be easily detectable. In double-stranded DNA condensation
experiments, the signature of the presence of rod-like structures was not
unambiguously detected, suggesting that the polyamines used to condense DNA may
protect it from bending sharply as needed in the rod-like structures.

\end{abstract}

\maketitle

\section*{Copyright notice}

Copyright (2015) American Institute of Physics.
This article may be downloaded for personal use only. Any other use requires
prior permission of the author and the American Institute of Physics.

The following article appeared in R.~Cortini \textit{et al.}, The Journal of
Chemical Physics \textbf{142}, 105102 (2015) and may be found at
\verb+http://dx.doi.org/10.1063/1.4914513+

% Introduction
\section{Introduction}
Single-molecule DNA micromanipulations provide a very powerful tool for the
study of DNA mechanics\cite{Smith1992,Strick1996}, DNA-DNA
interactions\cite{Murayama2003,Todd2008a,Todd2008b,vandenBroek2010}, and DNA-protein
interactions\cite{Noom2007}. In the case of DNA-DNA interactions, it was
possible to study the reentrant behavior of cation-condensed
DNA\cite{Murayama2003,Besteman2007a,Besteman2007b}, and measure accurately the
free energy of DNA condensation\cite{Todd2008b}. These studies are of great importance
to determine the behavior of condensed DNA, which is crucial for DNA packaging
and genomic ejection from viruses\cite{Carrivain2012}, and also to provide an
accurate estimate of the forces involved.

In single-molecule DNA condensation experiments, a DNA molecule is tethered at
one end to a surface, and at the other end to a micron-sized bead, which is
trapped by magnetic or optical tweezers. The experimental extension-force
curves show that there is a critical traction force, above which the system is
fully extended and behaves as a worm-like chain semiflexible polymer, and below
which the system progressively folds to a completely condensed
state\cite{Baumann2000,Murayama2003,Todd2008a,Todd2008b}. This critical
condensation force allows to estimate the attractive DNA-DNA interaction per
unit length. So far, this was done by neglecting finite size effects for long
$\lambda$ DNA molecules.

DNA condensation in single-molecule experiments was already studied theoretically
\cite{Battle2009,Cardenas2009}, but the geometry of the
condensed structure was always assumed \emph{a priori} to be toroidal. It is
possible however that toroids are not the only possible geometry for the
condensate, as was already noticed in theoretical studies of condensation of
semiflexible polymers with self-interactions\cite{Stevens2001,Sakaue2002}.
It is therefore interesting to explore the possibility of the appearance of other
geometrical configurations in single-molecule DNA condensation experiments.

In this paper we propose to study DNA condensation in single-molecule
experiments by means of Langevin dynamics (LD) simulations, together with an
analytical model. The scope of this work is two-fold: on the one hand, to
better account for finite-size effects in the estimate of attractive DNA-DNA
interactions using measured critical condensation forces; on the other hand, to
evaluate the possibility of the appearance of the rod-like geometry of the
condensate, alongside the well-known toroidal one.

Our LD simulation method is based on the work of Carrivain et
al\cite{Carrivain2014}, which has been successfully applied to modeling
single-molecule DNA supercoiling experiments\cite{Mosconi2009}. The main
advantage of using this technique is its speed and enhanced sampling
efficiency, based on (a) efficient algorithms to simulate rigid
bodies\cite{Smith2008} and (b) the global thermostat scheme introduced by Bussi
and Parrinello\cite{Bussi2008}. Using this method, we are able to observe
nucleation of DNA condensation within our simulation windows.  The well-known
toroidal structure of DNA condensates\cite{Hud2005} is not the only geometry
that we observe: rod-like structures also appear.

The analytical model we propose is based on the theoretical ansatz proposed by
Hoang et al.\cite{Hoang2014}. This was developed to study toroidal and rod-like
structures of free DNA in solution, during crowding- or cation-induced
condensation. We extended the model proposed by these authors to include the
presence of a traction force applied at the extremities of the DNA. Using
this model, we are able to evaluate the different contributions to the free
energy of the system: the bending energy, the surface tension, the bulk
attractive force, and the worm-like chain entropic free energy of the
non-condensed region of the DNA. We are also able to calculate the critical
condensation force at any DNA length.

In both our simulations and our analytical model we consistently model the
counterions that are present in solution implicitly, using a pairwise attractive
potential. The validity of this approximation is supported by a favorable
comparison of our results with experimental data on DNA condensation.

% Methods
\section{Methods}

\subsection{Geometry}
\begin{figure}[htp]
  \includegraphics[width=0.5\textwidth]{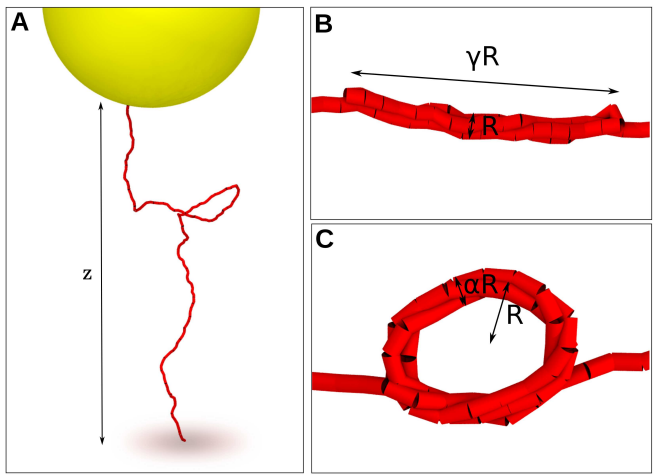}
  \caption{Geometry of the single-molecule DNA manipulation experiment.(A) Bead and
DNA molecule attached to it; (B) rod-like condensed structure: $R$ is the rod
diameter, $\gamma$ is the aspect ratio of the rod; (C) toroid-like
condensed structure: $R$ is the toroid radius, $\alpha$ is the ratio between the
radius and the thickness of the toroid. \label{fig:geometry}}
\end{figure}

The geometry of our system is described in figure \ref{fig:geometry}.
We describe a polymer of length $L$, tethered at one end to a
surface and at another end to a bead. In magnetic tweezer experiments, the force
$F$ acting on the bead is imposed, and the end-to-end distance $z$ is measured.
In optical tweezers, the bead is trapped around a fixed height $z_0$, and the
force is measured. These two experimental setups correspond to different
statistical ensembles: at fixed force (where we use the Gibbs free energy), and
at fixed end-to-end extension (where the Helmholtz free energy is used).

\subsection{Analytical model}
We describe the polymer as a made of two different phases: a condensed phase
characterized by a length $L_c$ and an extended phase of length $L-L_c$. We
therefore have the following two expressions for the free energy, at fixed force
and at fixed end-to-end extension respectively:
\begin{equation}
  G = E_c + (L-L_c) g_{WLC} (F),
  \label{eq:Gibbs_free_energy}
\end{equation}
\begin{equation}
  A = E_c + (L-L_c) a_{WLC} \left(\frac{z}{L-L_c}\right),
  \label{eq:Helmholtz_free_energy}
\end{equation}
where $E_c$ is the energy of the condensed phase, $g_{WLC}$ and $a_{WLC}$ are
the free energies per unit length of the worm-like chain phase, in the fixed
force and fixed extension ensembles, respectively. In our model, we neglect the
entropic contribution due to fluctuations inside the condensed phase: we
consider that the condensed phase has a definite conformational structure. We
now discuss separately the contributions from the condensed phase and from the
extended phase.

For the condensed phase, we base our model on the recent work by Hoang et
al.\cite{Hoang2014}. Here, we report the main results of this study. More
details on the derivation of the following formulas, as well as a discussion on
their validity, is reported in the original study\cite{Hoang2014}.

The authors model a polymer chain with self-interactions as
a chain of spheres of radius $b$ connected by rigid bonds, and propose an energy
ansatz for the toroidal and rod-like geometry. In our model, we take their
expressions, and assign the energy of the condensed phase $E_c$ to one of the
two formulas:
\begin{multline}
  E_{toroid} (\eta,\alpha, L_c) = 2 l_p \pi^{2/3} \eta^{2/3} \alpha^{4/3}
  b^{-4/3} L_c^{1/3} + \\ - \phi (d) d^{-1} b^{1/3} \pi^{4/3} \alpha^{-1/3}
  \eta^{-2/3} L_c^{2/3} + \\ + 3 \phi (d) b^{-1} L_c \label{eq:Etoroid}
\end{multline}
\begin{multline}
  E_{rod} (\eta,\gamma, L_c) = \frac{32}{3} l_p \eta \left[4\eta \left(\gamma +
  4/3\right)\right]^{-1/3} L_c^{1/3} + \\
  - 2 \pi \phi (d) d^{-1} b^{1/3} \left(\gamma+2\right) \left[4\eta \left(\gamma
  + 4/3\right)\right]^{-2/3} L_c^{2/3} + \\
  + 3 \phi (d) b^{-1} L_c.
  \label{eq:Erod}
\end{multline}
Both the toroid and the rod geometries are characterized by $\eta$, the packing
fraction in the condensate. The toroid is described also by $\alpha$, which is
the ratio between the toroid radius and the thickness of the toroid; the rod is
described by $\gamma$, the rod aspect ratio between the width and the length of
the rod (see figure \ref{fig:geometry}). The interaction of the spheres is given
by the potential function $\phi (d)$. We chose the Lennard-Jones potential
energy function, which is numerically very close to the more realistic Morse
potential:
\begin{equation}
  \phi (d) = \epsilon
  \left[\left(\frac{\sigma}{d}\right)^{12}-2\left(\frac{\sigma}{d}\right)^{6}\right].
  \label{eq:LJpotential}
\end{equation}
Here, $\sigma$ is the Lennard-Jones radius of the monomers. The variable $d$,
which represents the inter-monomer lateral distance in the condensed phase, is
related to $\eta$ by the following expression:
\begin{equation}
  d = b \sqrt{\eta_{hex}/\eta},
  \label{eq:d_eta}
\end{equation}
where $\eta_{hex} = \pi/6 \cot (\pi/6)$. Finally, $l_p$ is the polymer bending
persistence length. Notice that the precise form of the
interaction potential was found to be of minor importance in determining the
position and length dependence of the toroid/rod transition line in the work of
Hoang et al\cite{Hoang2014}. We also checked that the Morse potential gave
little or no difference compared to the the Lennard-Jones potential in our
analysis.

Both the toroid and the rod energy functions are a sum of three terms: the
bending energy to fold the polymer into its condensed state (which is $\sim
L_c^{1/3}$); the surface tension (which is $\sim L_c^{2/3}$), and the bulk
attractive energy, which is proportional to $L_c$. The surface tension and the
bulk energy terms are proportional to the amplitude of the attractive potential,
$\varepsilon$. These terms are derived by supposing that the packing of the
polymer in the condensate is hexagonal, so that one can calculate the average
number of neighbors in the bulk, and the average number of molecules at the
surface. As we shall see, the dependence of these terms on $L_c$ has important
consequences on the equilibrium properties of the system.

The free energy of the extended phase may be expressed using the formula derived
by Marko and Siggia\cite{Marko1995a} for the Gibbs free
energy per unit length of a worm-like chain semi-flexible polymer of length
$L-L_c$:
\begin{equation}
  g_{WLC} (F) = \min_a \left\{ \left( \frac{a}{2 l_p} - F\right) \left( \coth 2a -
  \frac{1}{2a}\right)\right\}.
  \label{eq:gWLC}
\end{equation}
This expression is used to calculate the relative extension $\rho = z/(L-L_c)$ at
thermal equilibrium
\begin{equation}
  \rho_{WLC} (F) = -\frac{\partial g_{WLC}}{\partial F}.
  \label{eq:rho}
\end{equation}

To express the free energy per unit length in the fixed end-to-end
extension ensemble, we must perform the Legendre transform of $g_{WLC}$:
\begin{equation}
  \tilde{a}_{WLC} = g_{WLC} (F) - F \rho.
  \label{eq:atilde_WLC}
\end{equation}
To obtain the force as a function of the extension, we invert the $\rho_{WLC} (F)$
function and we obtain the $F_{WLC} (\rho)$ function. Once equation
\ref{eq:atilde_WLC} is specified for $F = F_{WLC} (\rho)$, we finally obtain
\begin{equation}
  a_{WLC} = g_{WLC} \left[F_{WLC} (\rho)\right] - F_{WLC} (\rho) \rho.
  \label{eq:aWLC}
\end{equation}

The equilibrium condition is obtained by minimizing the free energy (either
equation \ref{eq:Gibbs_free_energy} or \ref{eq:Helmholtz_free_energy}) with
respect to all the variables.

We may derive interesting predictions from minimization of the Helmholtz free
energy (equation \ref{eq:Helmholtz_free_energy}). The minimum condition for $L_c$ is
given by
\begin{equation}
  \left.\frac{\partial E_c}{\partial L_c}\right|_{L_c = L_c^\star} - g_{WLC}
  \left[F_{WLC} \left(\frac{z}{L-L_c^\star}\right)\right] = 0,
  \label{eq:Lc_star}
\end{equation}
where the starred symbols refer to the equilibrium condition. The above equation is
derived in the Appendix. In the thermodynamic limit, $L_c \to \infty$, and
equation \ref{eq:Lc_star} may be approximated by
\begin{equation}
  -3 \varepsilon = g_{WLC} \left[F_{WLC} \left(\rho^\star\right)\right].
  \label{eq:Lc_star_approx}
\end{equation}
This equation depends only on $\rho^\star = z/(L-L_c^\star)$, and not on $z$ 
and $L$ separately. Therefore, in the limit of large $L$ we expect
$z/(L-L_c^\star)$ to be constant. This may be translated into the following
useful formula, which we will use later:
\begin{equation}
  L_c^\star = L - \frac{z}{\rho^\star}
  \label{eq:Lc_star_rho_star}
\end{equation}

At fixed end-to-end extension, the force acting on the bead at thermal equilibrium
is given by:
\begin{equation}
  F = F_{WLC} \left[\rho^\star\right].
  \label{eq:F_fixed_z}
\end{equation}
This equation is derived in the Appendix.

%We calculate the worm-like chain functions using the \verb+wlc+ package (\ldots).
To obtain the equilibrium structures, we use numerical routines for
minimization of multi-variable functions, as implemented in the GNU Scientific
Library\cite{GSL}.

To calculate the critical condensation force, we switch back to the fixed force
ensemble. The transition from the fully extended state to the fully
condensed state is obtained by equating the chemical potentials of these two
phases, at the critical force. We then obtain the following expression:
\begin{equation}
  \frac{E_c^\star}{L} = g_{WLC} (F_c),
  \label{eq:Fc}
\end{equation}
where in $E_c^\star$ we set $L_c = L$. This equation may be solved numerically
to obtain $F_c$. Since the bending energy and surface
tension terms vary as $\sim L_c^{1/3}$ and $\sim L_c^{2/3}$ respectively, in the
thermodynamic limit of $L \to \infty$, we obtain
\begin{equation}
  -3 \varepsilon = g_{WLC} (F_c^\infty),
  \label{eq:Fc_infinity}
\end{equation}
where $F_c^\infty$ is the critical force for an infinitely long chain.

\subsection{Molecular dynamics simulations}
Our simulation assay is based on the recently published work by Carrivain et
al\cite{Carrivain2014}. We mention here only the most relevant features of this
simulation method. First, it is a simulation of rigid bodies, where the angular
velocity is taken into account, and the simulated bodies have an impenetrable
volume. In our case, the DNA is composed by $N$ cylinders of length $l_s$,
with a hard-core radius equal to the DNA crystallographic radius, and is
attached to a magnetical or optical bead, which is treated as a sphere (see
figure \ref{fig:geometry}A). Second, the DNA chain is modelled by an
articulated system, where the joints act as holonomic contraints of a mechanical
system.  Forces and torques may be applied at each joint, so to model bending
and twisting energies.  In our case, we apply only a bending term, and neglect
the twisting contribution, which is irrelevant because the DNA molecules are
typically nicked in the experiments. Finally, the coupling to the thermal bath
is treated using the global thermostat scheme, introduced by Bussi and
Parrinello\cite{Bussi2008}. This allows for fast equilibration of the system and
realistic treatment of thermal fluctuations.

We model the bending energy associated to two successive cylinders as:
\begin{equation}
  E_b (\theta) = \frac{1}{2} g_b \theta^2,
  \label{eq:Ebending}
\end{equation}
where $\theta$ is the angle formed by the tangent vectors of the two 
cylinders, and $g_b$ is a constant related to the bending persistence length
(for full details, see reference \cite{Carrivain2014}).

Here, we include the effect of intra-DNA interactions, as modelled by the
Lennard-Jones potential equation \ref{eq:LJpotential}. In our system, the potential
acts between the centers of mass of the cylinders that compose the articulated
system. The Lennard-Jones radius $\sigma$ was chosen so that the minimum of the
potential corresponds to 28\AA{}. The full list of parameters used in our
simulations is given in table \ref{tab:parameters}.

To simulate optical tweezer experiments, we add a potential energy term to the
bead, expressed as an isotropic harmonic trap:
\begin{equation}
  E_{trap} = \frac{1}{2} k_{trap} (\left|\bm{r}-\bm{r}_0\right|)^2.
  \label{eq:opticaltrap}
\end{equation}
Here, $k_{trap}$ is the stiffness of the optical trap. Typical values of this
parameter are of the order of 0.1pN$/$nm\cite{Murayama2003}. The vectors
$\bm{r}$ and $\bm{r}_0 \equiv (x_0, y_0, z_0)$ are the actual position of the
bead, and the position of the center of the force (which would correspond to the
laser focus), respectively.

We run an equilibration round of 10$^6$ steps, using a local Langevin thermostat
and no inter-monomer potential. After this, we switch to the global thermostat
and turn on the inter-monomer interactions, and perform 10$^7$ LD steps for
production run.

\subsection{Data analysis}
We briefly describe here a few methods we used to analyze the data from our LD
simulations.

\noindent\textbf{Condensate length.} To calculate the length of the condensed
region from an LD trajectory, we proceed as follows. First, we calculate the
distance matrix between all the DNA segments at a given time step. That is,
calculate $M_{ij} = |\bm{r}_i - \bm{r}_j|$. Then we select all segments in which
$|i-j|>1$, and $M_{ij} < l_s$. The number of segments that satisfy this
condition, multiplied by $l_s$, gives $L_c$.

\noindent\textbf{Free energy.} The free energy of a given state may be estimated
by calculating the total energy of the condensate (Lennard-Jones potential plus
bending energy), and adding the worm-like chain contribution of the uncondensed
region. This can be calculated by averaging $L_c$ over the trajectory, and using
the expression for the worm-like chain free energy per unit length proposed by
Marko and Siggia (see equations \ref{eq:aWLC} and \ref{eq:gWLC}). As we have
done in our analytical model, we neglect the entropic contribution of
fluctuations of the segments in the condensed phase.

\noindent\textbf{Critical force.} It is challenging to estimate $F_c$ with LD
simulations directly. At the critical point, the time scale for the formation of
an initial condensation loop diverges. Therefore, it is difficult to estimate
$F_c$ by starting from a high force value and decreasing it, because the time
scale for nucleation is inaccessible in LD simulations. On the other hand, a
similar problem is encountered when starting from a condensed structure and
increasing progressively the applied force. Here, the kinetic barrier between
the folded and unfolded states is significant, therefore presenting the same
conceptual problem. However, one can estimate $F_c$ by fitting the
$\left<L_c\right> (\left<z\right>)$ curves to a line, and estimating the
intercept of the line with the $L_c = 0$ axis (see equation
\ref{eq:Lc_star_rho_star} and figure \ref{fig:Lc_rho}). The value of $z$ at
which $L_c = 0$ gives an estimate of $\rho^\star$, from which we can estimate
the critical force using the $F_{WLC} (\rho)$ function.

\noindent\textbf{Distinguishing between toroids and rods.} To automatically
detect whether a condensed state is toroidal or rod-like, we developed an
heuristic method outlined below. Once the cylinders that are part of the
condensed phase are identified, for each pair of cylinders $i$ and $j$ in
contact, we evaluate $\cos \gamma_{ij} = \bm{t}_i \cdot \bm{t}_j$.  If $\cos
\gamma_{ij} > 0$, we say that those segments are in parallel contact, otherwise
they are in antiparallel contact. We can then define an order parameter as $s =
(N_p - N_a)/N_c$, where $N_p$ and $N_a$ are the number of parallel and
antiparallel contacts, respectively, and $N_c$ is the total number of contacts.
For a toroid, most segments will be in parallel contact, so we expect $s \approx
1$. On the other hand, for a rod-like configuration we expect $s \approx 0$,
since $N_p \approx N_a$.

% Results
\section{Results}
We performed Langevin dynamics simulations of a 3 kb DNA molecule ($N = 150$,
$L = 1\mu m$), as described in Methods, using parameters shown in table
\ref{tab:parameters}, in an optical tweezer-like setup. We compared the
simulation results with the predictions of our analytical model (equations
\ref{eq:Gibbs_free_energy} and \ref{eq:Helmholtz_free_energy})
using the same parameters, for the toroidal and rod-like geometries. We present
here the results of these two approaches.
\begin{table*}[htp]
  \centering
  \begin{tabular}{l l l }
  Parameter & Description & Value \\
  \hline
  \hline
  $l_p$   & Bending persistence length & 45 nm \\
  $T$   & Absolute temperature & 300 K \\
  $N$   & Number of DNA segments & 150 \\
  $n_s$   & Number of base pairs per DNA segment & 20 \\
  $\Omega$   & Thermostat coupling frequency & 3 $10^{10} s^{-1}$ \\
  $k_{trap}$   & Optical trap stiffness & 0.2 pN nm \\
  $\varepsilon$   & Lennard-Jones potential well depth & 0.2--0.7 $\kt/\mathrm{nm}$ \\
  $\sigma$   & Lennard-Jones radius & 25.08 \AA{} \\
  \hline
  \end{tabular}
  \caption{Summary of the parameters used in Langevin dynamics simulations and
    in our analytical model (where applicable).
  \label{tab:parameters}}
\end{table*}

\subsection{Fixed $z_0/L$ simulations}
First, we performed 20 independent simulations of the DNA chain, at fixed
$z_0/L = 0.6$ and at $\varepsilon = 0.7 \kt/\mathrm{nm}$. We observed that after
$10^7$ time steps, 9 chains adopted the rod-like condensed state, 10 adopted the
toroidal state, and one remained uncondensed. Snapshots of the rod-like and
toroidal states are shown in figure \ref{fig:geometry}B and \ref{fig:geometry}C.

\begin{figure}[htp]
  \centering
  \includegraphics[width=0.5\textwidth]{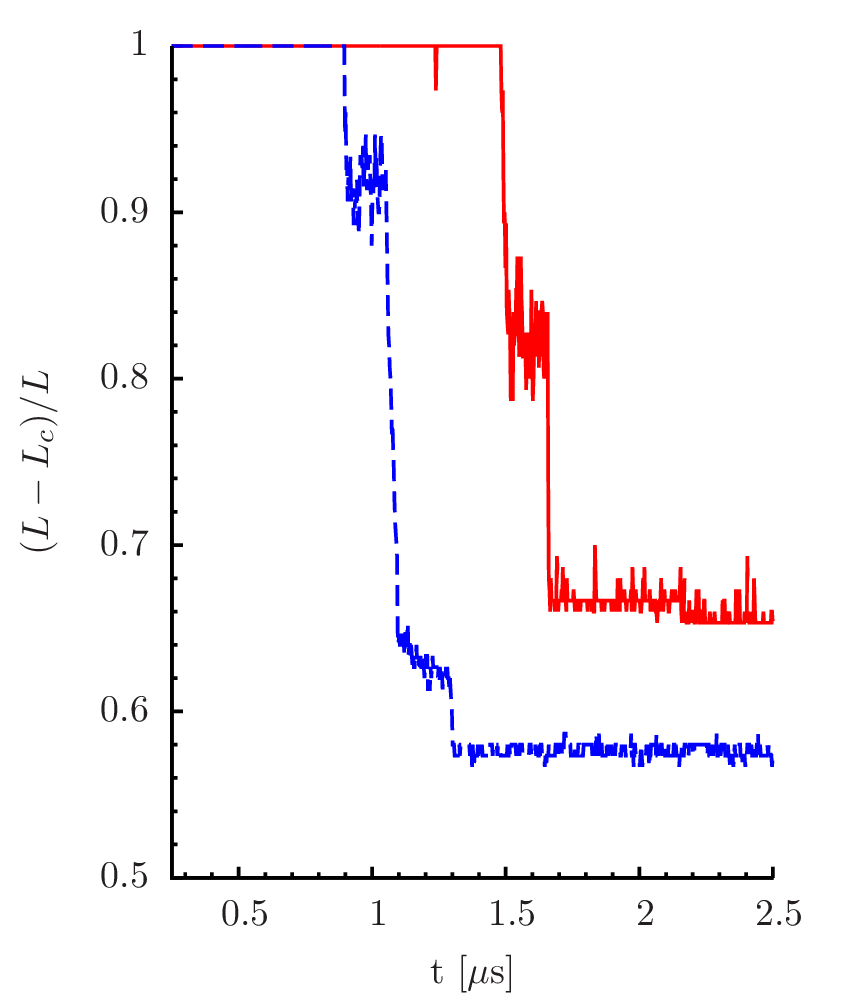}
  \caption{Variation of the uncondensed length $(L-L_c)/L$ ($L_c$ is calculated
  as explained in Methods, as a function of time for two specific simulations,
one which nucleates into a rod-like condensate (solid red) and one which
nucleates into a toroid (dashed blue).\label{fig:Lc_t}}
\end{figure}

Figure \ref{fig:Lc_t} shows two example traces of $(L-L_c (t))/L$ for a toroid
and a rod. The condensation of the chain occurs as discrete jumps, both for the
toroid and rod-like geometries. The steps correspond to adhesion of successive
portions of the DNA chain onto an initially formed condensation loop for toroids
and a condensation stretch for rods. In the time trace of the toroid
condensation, a more progressive, linear condensation step is also observed.

\subsection{Force-extension curves}
\begin{figure}[htp]
  \centering
  \includegraphics[width=0.5\textwidth]{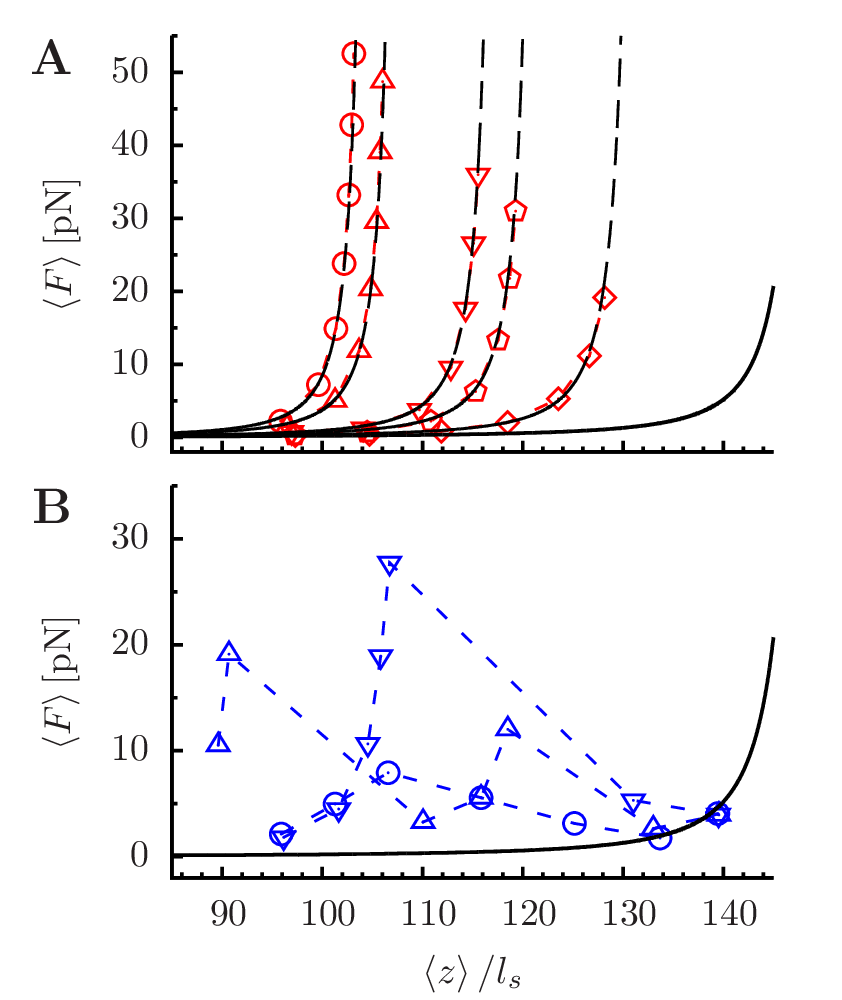}
  \caption{Average force versus average extension of single stretched rods (A)
    and toroids (B). The different symbols correspond to simulations starting
    from independent configurations (see Main Text). The worm-like chain
    behavior for a completely uncondensed DNA is shown as a black continuous
    line. In A, black dashed lines correspond to the fit to a worm-like chain
    behavior for, from left to right, $L/l_s = 105.5$, $L/l_s = 108.5, 118.5,
    122.5, 132.5$.  The simulation protocol is described in Methods, and the
    parameters used are shown in table \ref{tab:parameters}.
\label{fig:force_extension}}
\end{figure}

Next, we used the final configurations of the simulations described in the
previous section to perform stretching simulations. Here, every $10^6$ time
steps the value of $z_0/L$ was incremented by 5\%. The resulting force-extension
curves are depicted in figure \ref{fig:force_extension}.
The toroid- and rod-like structures have a markedly different response under
stretching force. The toroidal structures unfold progressively, exhibiting a
saw-tooth-like force-extension behavior, eventually becoming fully extended (see
final blue points in figure \ref{fig:force_extension}). The peaks in the force
value can go up to $\sim$30pN. In contrast, rod-like structures never unfold,
even at the highest forces probed here ($\sim$50pN). In all our simulations, we
could not unfold rod-like structures, whereas toroidal structures all eventually
unfold.

The difference in the force-extension behavior of toroids and rods under tension
can be understood in terms of the direction of the applied force. In a rod-like
condensate, the rod axis is in general approximately parallel
to the direction of the applied force (see figure \ref{fig:geometry}B). As the
force increases, the direction of the force becomes more and more orthogonal to
the direction of the DNA-DNA contacts. This makes it difficult for the force to
unfold the condensate. On the other hand, the toroidal structures have a larger
degree of lateral mobility, that results in fluctuations aligning the lateral
contacts with the direction of the applied force.

\subsection{Free energy of toroidal and rod-like structures}
\begin{figure}[htp]
  \centering
  \includegraphics[width=0.5\textwidth]{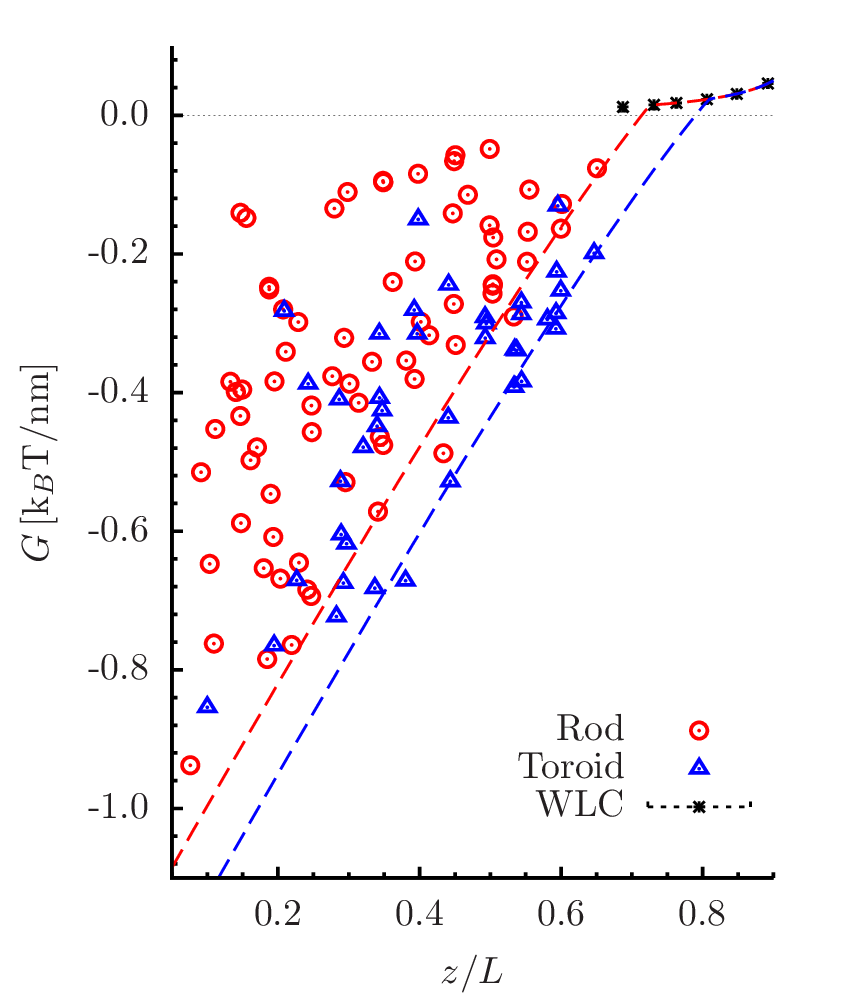}
  \caption{Minimum free energy of rod-like and toroidal geometries (blue and
    red dashed lines, respectively), as calculated from numerical minimization
    of equation \ref{eq:Helmholtz_free_energy}, and average
free energies of final states in Langevin dynamics simulations, calculated as
explained in Methods, for rods (red circles) and toroids (blue triangles).
\label{fig:free_energy_rho}}
\end{figure}

To further gain insight on the system, and to test the results of our analytical
model (equations \ref{eq:Gibbs_free_energy} and \ref{eq:Helmholtz_free_energy}),
we performed simulations at different values of $z_0/L$. For each value of
$z_0/L$, we obtained results for 10 independent starting configurations. We then
calculated the free energy of the final configurations as outlined in Methods,
averaging over the 10 simulations. Figure \ref{fig:free_energy_rho} shows the
comparison of the numerical minimization of our analytical model, for the toroid
and rod geometries, and the free energies obtained from Langevin dynamics
simulations as outlined above.

For each value of $z_0/L$ that we studied, the toroidal states are the ones
having the lowest free energy. The free energy minima for toroidal and rod-like
configurations, however, are remarkably close (within $\sim 0.1 \kt/\mathrm{nm}$
difference). This fact is also manifest in the observation of simulations in
which a toroidal and a rod-like condensed phase coexist within the same DNA
chain (not shown).

Our simulations show that at $\left<z\right>/L>0.7$ for toroids and
$\left<z\right>/L>0.6$ for rods, the condensed state ceases to exist.  At values
of $\left<z\right>/L$ higher than this, the free energy is equal to the
worm-like chain free energy. 

We notice that there is a significant spread between the values of the free
energies calculated in our simulations. However, the lowest values of the free
energy for a given conformation are remarkably close to the predictions of our
analytical model. This means that in the other cases, the conformation adopted
by the chain is not an equilibrium one, but a local minimum. This is a sign that
the final conformation is strongly dependent on the nucleation of a first
condensation loop.

\subsection{Critical condensation force}
\begin{figure}[htp]
  \centering
  \includegraphics[width=0.5\textwidth]{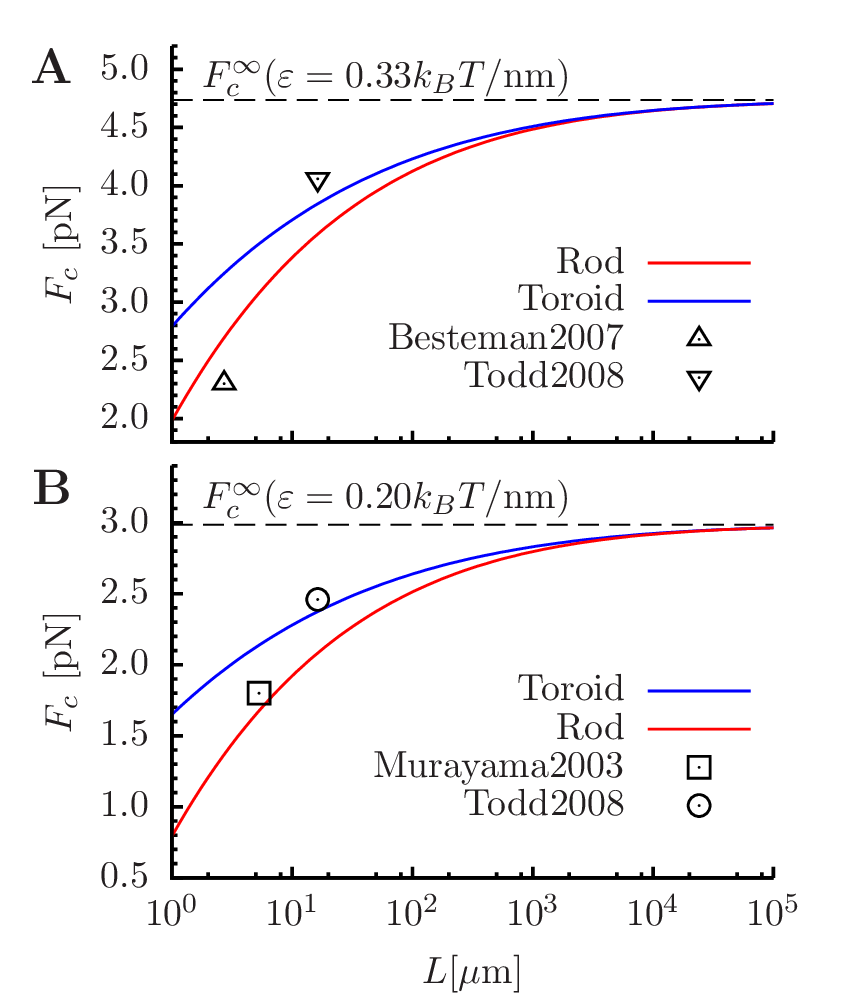}
  \caption{Dependence of the critical condensation force $F_c$ on the length of
  the DNA, for $\varepsilon = 0.33 \kt/\mathrm{nm}$ (A) and $\varepsilon =
  0.20 \kt/\mathrm{nm}$ (B). These two values correspond to the case of
  spermidine and spermine, respectively. The points are extracted from the experimental
  data of Murayama et al\cite{Murayama2003}, Besteman et al\cite{Besteman2007b},
  and Todd et al\cite{Todd2008a}.\label{fig:Fc_L_eps}}
\end{figure}

Figure \ref{fig:Fc_L_eps} shows the values of the critical condensation force
$F_c$ as estimated from the analytical model (see equation \ref{eq:Fc}), as a
function of the DNA length, at fixed $\varepsilon$. The two values chosen for
$\varepsilon$ correspond to the case of spermidine (0.20 $\kt/\mathrm{nm}$) and
spermine (0.33 $\kt/\mathrm{nm}$) given in reference \cite{Todd2008a}. As we
explained in the Methods section, for a very long chain the critical
condensation force reaches a finite value $F_c^\infty$, and becomes independent
of the geometrical structure of the condensate. Here, we show that this
thermodynamic limit is reached only for very long molecules. As an example, for
$\varepsilon = 0.33 \kt/\mathrm{nm}$, the value of $F_c$ reaches $F_c^\infty$
only for $L = 10^8\mathrm{nm}$, the size of a genomic DNA molecule. The
experimental values of the maximum condensation force, as extracted from the
work of Todd et al\cite{Todd2008a} and Murayama et al\cite{Murayama2003} are
also depicted. There is good agreement between the theoretical and experimental
values.

The reason for the very slow convergence of $F_c$ to $F_c^\infty$ is to be found
in the surface tension term. In fact, this term is proportional to $\varepsilon$
(see equations \ref{eq:Etoroid} and \ref{eq:Erod}, which is typically a very
significant value, and depends on $L_c^{2/3}$, which is not far from linear
dependence. Therefore, this term represents a significant energy penalty even
for a very long DNA.

\begin{figure}[htp]
  \centering
  \includegraphics[width=0.5\textwidth]{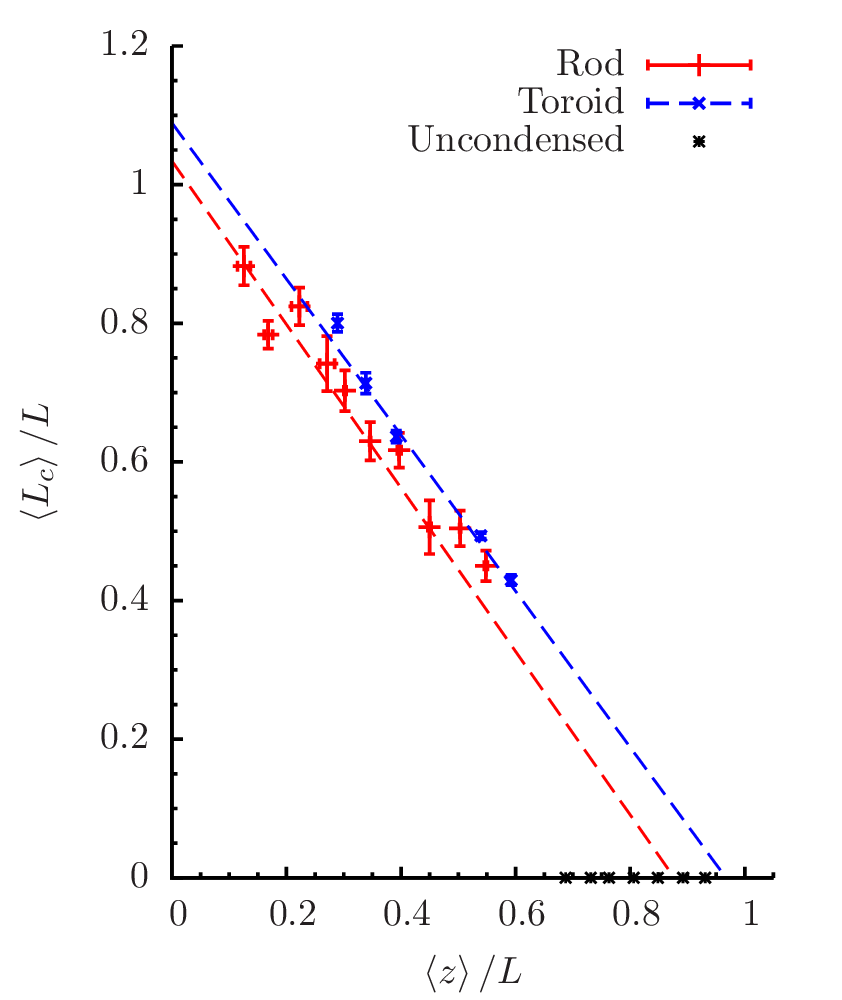}
  \caption{Condensed length in simulations, as a function of average
    $\left<z\right>/L$. Condensed length was calculated as described in Methods.
    Error bars are standard deviations. The  dashed lines are linear fits to the
  data.\label{fig:Lc_rho}}
\end{figure}

As outlined in the Methods section, one can estimate the critical condensation
force from LD simulations by looking at the dependence of the condensed length $L_c$ 
on $z$. Figure \ref{fig:Lc_rho} shows the variation of the average $L_c$, for
toroids and rods, at fixed $z$. The data clearly shows a linear dependence of
$<L_c>$ on $<z>$, which is one of the key predictions of our analytical model
(see Methods).
A linear fit of this data provides a value of the intercept with the $z = 0$
axis which is close to 1, meaning that in the absence of an external force, the
whole molecule is in the condensed state. The value of $\rho^\star$ estimated
from our data is 0.96 and 0.87 for toroids and rods, respectively.
Unfortunately, small errors in $\rho^\star$ result in large errors on the
estimate of $F_c$, as for $\rho > 0.85$ the $F_{WLC} (\rho)$ function has a very
steep increase (see figure \ref{fig:force_extension} for comparison). Therefore,
we estimate $F_c$ to be between $\approx$1.5 and 20pN, at $\varepsilon = 0.7
\kt/\mathrm{nm}$.

% Discussion
\section{Discussion}
In this study, we developed an analytical theory and a simulation method to
investigate the behavior of a DNA under traction and in the presence of
self-attraction. This study was primarily aimed at the investigation
of single-molecule DNA condensation in optical and magnetic tweezers
\cite{Baumann2000,Murayama2003,Fu2006,Besteman2007b,Todd2008a,Todd2008b,vandenBroek2010}.
Whereas our simulations allow to study the
kinetic aspects of DNA condensation, our analytical theory is aimed at the study
of the equilibrium properties of the system. The latter are experimentally
accessible, using a very slow unloading rate of a magnetic bead (see
e.~g.~ Ref.~\cite{vandenBroek2010}). In this section we discuss our results in
relation to the experimental data.

\subsection{General features of single-molecule DNA condensation experiments
are reproduced by our simulations}
Firstly, we notice that several experimental features of DNA condensation are
well captured by our model. The condensation of the chain proceeds in discrete
steps, and each step was shown to correspond to folding of a portion of the
molecule onto an initial condensation loop\cite{Fu2006,vandenBroek2010}. When
the condensate is toroidal, the force-extension curve has the characteristic
saw-tooth shape, which was seen in experiments\cite{Murayama2003} and in
previous simulations of DNA condensation \cite{Cardenas2009}. The force peaks of
our simulations are quantitatively in agreement with the experimental ones, being
in the 10-30 pN range (see figure \ref{fig:force_extension}). Our values are
somewhat higher than the experimental ones, and this may due to the limited time
window accessible by LD simulations.  Moreover, we find that the conformation of
the condensed structures apprearing in the LD simulations is strongly dependent
on the size of the initial condensation loop, which is a well-known feature of
DNA toroids observed in electron microscopy
experiments\cite{Shen2000,Hud2001,Conwell2003}. 

Despite its simplicity, our simple assumption of a pairwise potential
between the monomers is able to capture some of the principal features of the
system.

\subsection{Evaluating the amplitude of DNA-DNA interactions using values of critical
condensation force}
\begin{figure}[htp]
  \centering
  \includegraphics[width=0.5\textwidth]{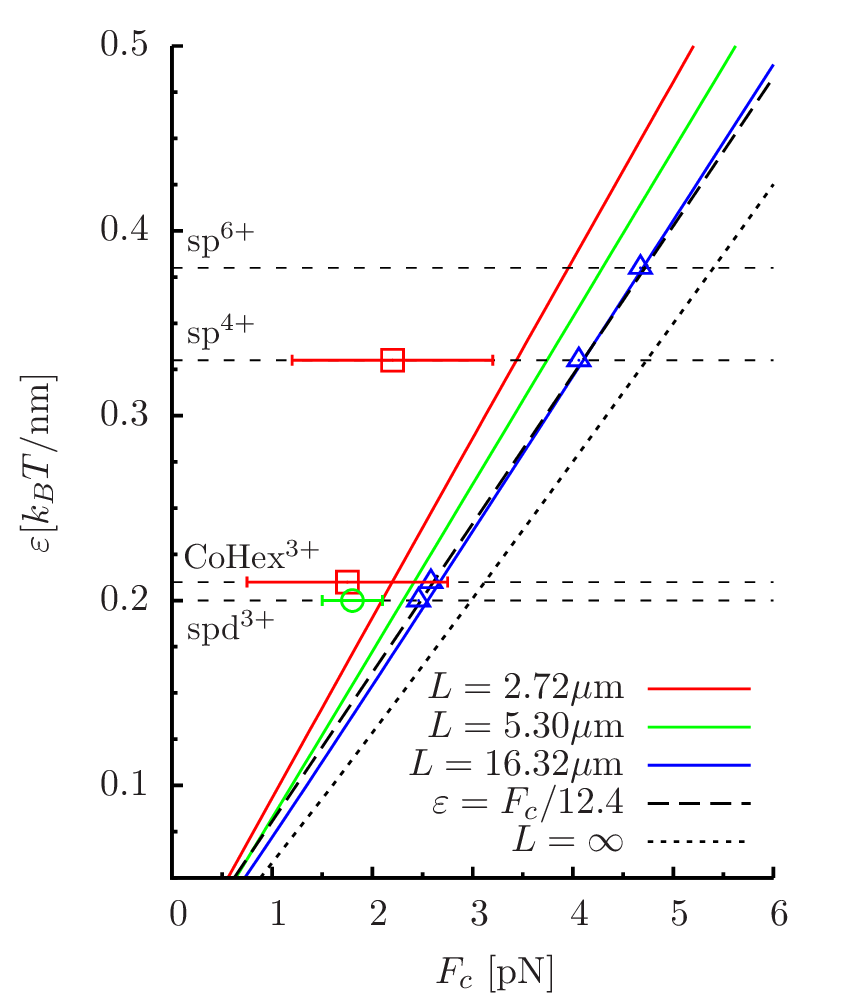}
  \caption{Estimated value of the amplitude of the DNA-DNA attraction
    $\varepsilon$, from a given measured value of the maximum critical
    condensation force $F_c$. Solid colored lines correspond to the theoretical
    prediction of our analytical model, as extracted by numerically solving
    equation \ref{eq:Fc}. Dotted line corresponds to the case of $L = \infty$.
    Dashed line corresponds to the estimate done by Todd et al\cite{Todd2008a},
    assuming direct proportionality between $\varepsilon$ and $F_c$.
    Experimental points were assigned by assuming that the values of
    $\varepsilon$ estimated by Todd et al\cite{Todd2008a}(blue triangles) were
    correct (see main text for further details) at a given ionic condition, and
    assigning the points extracted from the studies of Murayama et
  al\cite{Murayama2003}(green circle) and by Besteman et
al\cite{Besteman2007b}(red squares) to the corresponding ionic
conditions.\label{fig:eps_Fc}}
\end{figure}

The analysis of the dependence of the critical condensation force on the length
of the DNA gives an interesting prediction: the limit of infinitely long
chains is reached very slowly, as shown in figure \ref{fig:Fc_L_eps}. In the
work of Todd et al\cite{Todd2008a,Todd2008b}, the authors accurately measured
the condensation force at different ionic concentrations. The condensation force
has a maximum value as a function of the concentration of multivalent ions. For
this value, one can show that the entropic contribution to the free energy due
to ion mixing vanishes, and one can estimate the amplitude of the attractive
DNA-DNA interactions using the value of the maximum critical force. This
estimate is based on the idea that for a very long DNA molecule, the total free
energy of the condensate may be approximated by its bulk value. Our study
provides a quantitative way of veryfing this hypothesis. Figure \ref{fig:eps_Fc}
shows the relationship between the estimated values of $\varepsilon$ and the
condensation force. The dashed line corresponds to the hypothesis of Todd et al,
i.~e.~ assuming direct proportionality between $\varepsilon$ and $F_c$,
independently of the molecular length. Remarkably, the curve for $L = 16.3
\mu\mathrm{m}$ (corresponding to the $\lambda$ DNA molecules used in the
experiments of Todd et al) falls almost exactly on the dashed line.  Therefore,
even though $\varepsilon$ should have been estimated more rigorously, the values
supplied by Todd et al for $\varepsilon$ are correct.
Using the fact that those values are correct, we also set other values of
experimental critical forces in the figure, as extracted from the works of
Murayama et al\cite{Murayama2003} and Besteman et al\cite{Besteman2007b}, along
with the theoretical estimates corresponding to the molecular lengths used in
those experiments. The theoretical and experimental values are in reasonable
agreement.

\subsection{Force-extension curves allow to discriminate between toroids and
rods}
This study also showed that rod-like condensates have a different
force-extension behavior compared to toroids. In the paper by Baumann et al.
\cite{Baumann2000}, it was reported that some of the plasmid-length DNA
molecules could not be stretched to the full original extension. Our results may
provide a possible explanation for the two different force-extension behaviors
seen experimentally. To our knowledge, the results of Baumann et al. are the
only ones that show two different force-extension curve classes. In other
experimental works\cite{Murayama2003,vandenBroek2010} on the same system, this was not
reported.

The geometrical configuration of the condensate is highly dependent on the model
chosen for the DNA bending. If one allows for formation of kinks in the chain,
with low energetic cost, the rod-like structure becomes the most favorable geometry
for the condensate. In contrast, adding a high energy cost for the hairpin turns
strongly favors the toroidal geometry (data not shown). These results are in line
with previous theoretical studies\cite{Noguchi1998,Stevens2001}. This,
together with the fact that the force-extension signature of rods was not
clearly reported in the literature, suggests that the high bending angles
required in the rod-like structures have a higher energetic cost than that
expected by a simple harmonic model as we used here (see equation
\ref{eq:Ebending}). One possibility is that the polyamines used in DNA
condensation experiments protect the secondary structure of DNA from kinking or
bending sharply. It is in fact known that the polyamines stabilize
base-pairing\cite{Ouameur2004} and do not disrupt the B-DNA structural form
\cite{Deng2000}. It is therefore unlikely that DNA kinks are
favored by inhomogeneous patterns of adsorption of counterions on DNA.

The structure of condensed DNA is a subject of great
interest\cite{Carrivain2012}, because of the importance, for instance, of DNA
ejection from viral capsids\cite{Leforestier2009}. Our results show that under
the influence of an external traction force, the toroidal condensed structures
are the only ones that can readily unfold. When a DNA molecule becomes rod-like,
the unfolding is strongly impeded. This could be an important element to take
into account when envisaging therapeutic strategies against DNA viruses.

% Acknowledgments
\acknowledgments
The authors wish to thank D.~J.~Lee, A.~Kornyshev, A.~Giacometti, G.~Wuite, and
A.~Grosberg for useful discussions. This work has been funded by the French
Institut National du Cancer, grant INCa 5960 and by the French Agence Nationale
de la Recherche, grant ANR-13-BSV5-0010-03.

% Appendix
\appendix*
\section{Derivation of equations \ref{eq:Lc_star} and \ref{eq:F_fixed_z}}
The force acting on the bead at thermal equilibrium is obtained by taking the
total derivative of the minimum Helmholtz free energy with respect to $z$:
\begin{multline}
  F = \frac{\de A^\star}{\de z} = \left.\sum_i \frac{\partial E_c}{\partial
  X_i}\right|_{X_i = X_i^\star} + \\
  + \frac{\partial}{\partial z} \left.\left[ (L-L_c)
  a_{WLC} \left( \frac {z}{L-L_c}\right)\right]\right|_{L_c=L_c^\star},
  \label{eq:F}
\end{multline}
where the starred symbols indicate the equilibrium values of the variables. In
the above expression, the first term on the right-hand side vanishes, because
the gradient of the condensed energy is zero at equilibrium. The second term is
the only one that contributes to the force, and may be calculated by
straightforward differentiation. The first step is to calculate the derivative
of $a_{WLC}$ with respect to $\rho$:
\begin{multline}
  \frac{\partial}{\partial \rho} a_{WLC} (\rho)  = \frac{\partial}{\partial \rho}
  \left\{g_{WLC} \left[F_{WLC} (\rho)\right]
  + F_{WLC} (\rho) \rho\right\} = \\
  = \frac{\partial F_{WLC}}{\partial \rho} \frac{\partial g_{WLC}}{\partial F_{WLC}} +
\rho \frac{\partial F_{WLC}}{\partial \rho} + F_{WLC} (\rho) = \\
= F_{WLC} (\rho),
  \label{eq:daWLC}
\end{multline}
where we have taken into account equation \ref{eq:rho}. Using this result we may
now calculate easily the force acting on the molecule at equilibrium by
inserting equation \ref{eq:daWLC} into equation \ref{eq:F}:
\begin{equation}
  F = \frac{\partial}{\partial z} \left.\left[\left(L-L_c\right) a_{WLC}
\left(\frac{z}{L-L_c}\right)\right]\right|_{L_c=L_c^\star} = F_{WLC}
(\rho^\star),
  \label{eq:F1}
\end{equation}
which is equation \ref{eq:F_fixed_z}.

We may now explicitly calculate the derivative of the Helmholtz free energy
equation \ref{eq:Helmholtz_free_energy} with respect to $L_c$:
\begin{equation}
  \left.\frac{\partial E_c}{\partial L_c}\right|_{L_c = L_c^\star} -
  \frac{\partial}{\partial L_c} \left.\left[ \left(L-L_c\right) a_{WLC}\left(
  \frac{z}{L-L_c}\right)\right]\right|_{L_c=L_c^\star} = 0.
  \label{eq:dA_dLc}
\end{equation}
The second term on the left-hand side of this equation can be calculated using
again equation \ref{eq:daWLC}:
\begin{multline}
  \frac{\partial}{\partial L_c} \left[ \left(L-L_c\right) a_{WLC}\left(
  \frac{z}{L-L_c}\right)\right] = \\
  = -a_{WLC}(\rho) + (L-L_c) \frac{\partial\rho}{\partial L_c}
  \frac{\partial}{\partial \rho} a_{WLC}(\rho) = \\
  = -a_{WLC}(\rho) + (L-L_c) \frac{z}{(L-L_c)^2}F_{WLC} \left(\rho\right) = \\
  = -g_{WLC} \left[F_{WLC} \left(\frac{z}{L-L_c}\right)\right].
\end{multline}
Using this result, equation \ref{eq:dA_dLc} becomes
\begin{equation}
  \left.\frac{\partial E_c}{\partial L_c}\right|_{L_c = L_c^\star} - g_{WLC}
  \left[F_{WLC} \left(\rho^\star\right)\right] = 0,
  \label{eq:Lc_star_1}
\end{equation}
which is equation \ref{eq:Lc_star}.

% Bibliography
% \bibliographystyle{unsrtnat}
% \bibliography{Cortini_et_al_JCP_rev1}

\end{document}